\documentclass[aps,prc,twocolumn,superscriptaddress,nofootinbib]{revtex4-1}

\usepackage{amssymb}
\usepackage{amsmath}
\usepackage{siunitx}
\usepackage{pifont}
\usepackage{multirow}
\usepackage{color}

\usepackage{hyperref}
\hypersetup{letterpaper, colorlinks=true, breaklinks=True, citecolor=blue}

\usepackage{nicefrac}

\usepackage{graphicx}

\makeatletter

\newcommand{\Rmnum}[1]{\expandafter\@slowromancap\romannumeral #1@}
\makeatother

  \newcommand {\nc} {\newcommand}
  \nc {\beq} {\begin{eqnarray}}
  \nc {\eeq} {\nonumber \end{eqnarray}}
  \nc {\eeqn}[1] {\label {#1} \end{eqnarray}}
  \nc {\eol} {\nonumber \\}
  \nc {\eoln}[1] {\label {#1} \\}
  \nc {\ve} [1] {\mbox{\boldmath $#1$}}
  \nc {\ves} [1] {\mbox{\boldmath ${\scriptstyle #1}$}}
  \nc {\mrm} [1] {\mathrm{#1}}
  \nc {\half} {\mbox{$\frac{1}{2}$}}
  \nc {\thal} {\mbox{$\frac{3}{2}$}}
  \nc {\fial} {\mbox{$\frac{5}{2}$}}
  \nc {\la} {\mbox{$\langle$}}
  \nc {\ra} {\mbox{$\rangle$}}
  \nc {\etal} {\emph{et al.}\ }
  \nc {\eq} [1] {(\ref{#1})}
  \nc {\Eq} [1] {Eq.~(\ref{#1})}
  \nc {\Sec} [1] {Sec.~\ref{#1}}
  \nc {\chap} [1] {Chapter~\ref{#1}}
  \nc {\anx} [1] {Appendix~\ref{#1}}
  \nc {\tbl} [1] {Table~\ref{#1}}
  \nc {\Fig} [1] {Fig.~\ref{#1}}
  \nc {\ex} [1] {$^{#1}$}
  \nc {\Sch} {Schr\"odinger }
  \nc {\flim} [2] {\mathop{\longrightarrow}\limits_{{#1}\rightarrow{#2}}}
  \nc {\IR} [1]{\textcolor{red}{#1}}
  \nc {\IB} [1]{\textcolor{blue}{#1}}
  \nc{\IG}[1]{\textcolor{green}{#1}}

\begin{document}


\title{Study of the Coulomb and nuclear breakup of $^{11}$Be using an effective-potential description at N$^2$LO}


\author{L.-P. Kubushishi}
\email{lkubushi@uni-mainz.de}
\affiliation{Institut f\"ur Kernphysik, Johannes Gutenberg-Universit\"at Mainz, D-55099 Mainz, Germany}
\author{P. Capel}
\email{pcapel@uni-mainz.de}
\affiliation{Institut f\"ur Kernphysik, Johannes Gutenberg-Universit\"at Mainz, D-55099 Mainz, Germany}



\date{\today}
 
\begin{abstract}

\begin{description}
\item[Background] Halo effective field theory (Halo-EFT) provides a very efficient description of loosely-bound nuclei in models of reaction.
It offers a very systematical ranking of the significance of nuclear-structure observables in reaction calculations.
This greatly helps to infer reliable structure information from reaction cross sections.
However, for a meaningful analysis, the Halo-EFT scheme needs to have converged.
\item[Purpose] In a previous study [P. Capel, D. R. Phillips, and H.-W. Hammer, Phys. Rev. C {\bf 98}, 034610 (2018)], NLO descriptions of $^{11}$Be have been developed and lead to excellent agreement with existing breakup data.
However, the convergence of the scheme at NLO was not fully demonstrated.
Moreover, a significant dependence on the regulator of the effective $^{10}$Be-$n$ interaction has been observed.

\item[Method] We develop effective-potential descriptions of $^{11}$Be at N$^2$LO and use them in an accurate breakup-reaction code.
We compare our theoretical cross sections with experiment on Pb and C targets at about 70~MeV/nucleon.
\item[Results] On Pb, the N$^2$LO descriptions of $^{11}$Be lead to little change to the NLO results of the previous study, supporting the convergence of that scheme.
On C, the reaction is significantly affected by the presence of $d$ resonances in the low-energy spectrum of $^{11}$Be.
In effective-theory power counting these resonances appear only at N$^2$LO; our new descriptions include them naturally.
Going to N$^2$LO removes also the cutoff dependence observed in the previous study.
\item[Conclusions] We demonstrate the convergence of the effective description of $^{11}$Be at NLO for Coulomb breakup and at N$^2$LO for nuclear-dominated dissociation.
The reliability of the nuclear-structure information inferred in the previous study is thus confirmed.
\end{description}
\end{abstract}


\maketitle

\section{Introduction}\label{intro}

Halo nuclei exhibit one of the most peculiar quantal structures \cite{Tan96}.
They have a large matter radius compared to their isobars.
This unusual size arises from their very loose binding of one or two valence neutrons, which can thus tunnel far away from the other nucleons.
Accordingly, halo nuclei can be seen as a compact core, which contains most of the nucleons, surrounded by a diffuse neutron halo \cite{HJ87}.
This structure challenges modern nuclear models.
It is thus at the center of many experimental and theoretical studies \cite{Tan96}.

Because they are found close to the neutron dripline, halo nuclei are mostly studied through reactions \cite{AN03,BC12}.
To infer valuable information from experimental cross sections, a reliable model of the reaction coupled to a realistic description of the projectile is needed.
Halo effective field theory (Halo-EFT) provides a way to describe halo nuclei by exploiting the clear separation of scales observed in these nuclei.
The ratio of the small size of the core to the broad extension of the halo provides a naturally small parameter along which the core-halo Hamiltonian can be expanded \cite{BHvK02,BHvK03} (see Ref.~\cite{HJP17} for a recent review).
Extending this development at several orders, from the leading order (LO) to the next-to-leading order (NLO) and beyond, provides a systematic expansion scheme.
The low-energy constants (LECs) of this expansion are adjusted to physical observables taken either from experiment or from accurate nuclear-structure calculations.
This procedure therefore offers a natural scaling of the structure observables of halo nuclei.
Using a Halo-EFT description of the projectile in an accurate reaction code enables us to study the effect of these observables upon reaction cross sections and to confront the predictions of nuclear-structure models to experiment.
Originally considered for the breakup of $^{11}$Be, the archetypical one-neutron halo nucleus \cite{CPH18}, the idea has then been successfully extended to transfer \cite{YC18} and knockout \cite{HC21}, as well as to other halo nuclei: $^{15}$C \cite{MYC19,HC21} and $^{19}$C \cite{CP23}.

These various analyses have not only led to excellent agreement with experiment for various kinds of cross sections, but they also have enabled us to figure out accurately and systematically to which structure observables these reactions are sensitive.
In most cases, it seems that a NLO Halo-EFT description of the projectile is sufficient, which confirms that the reactions are mostly sensitive to the binding energy and asymptotic normalization constant (ANC) of the initial bound state as well as the phase shift in the continuum \cite{CN07}.
However, as pointed out during the program \emph{``Living Near Unitarity''} held at the KITP in Santa Barbara in 2022, the good agreement with experiment is not sufficient to conclude that the scheme has actually converged \cite{Gri22}.
Moreover, the ``beyond-NLO'' expansion developed in Ref.~\cite{CPH18} to include the $\fial^+$, $\thal^-$, and $\thal^+$ resonances in the description of $^{11}$Be exhibits a significant dependence on the regulator $\sigma$ of the effective $^{10}$Be-neutron interaction.
This is particularly true for the $p_{3/2}$ phase shift (see Fig.~10 of Ref.~\cite{CPH18}), which affects significantly the Coulomb-breakup cross section (see Fig.~12(a) of that reference).
This seeming dependence to the short-range physics of the projectile is neither physical \cite{CN07} nor expected within an EFT.

To address these two issues---lack of proof of convergence and $\sigma$ dependency of the description of $^{11}$Be---we follow Lepage \cite{Lepage97} and develop an effective-potential description of $^{11}$Be at N$^2$LO, and then repeat the reaction calculations performed in Ref.~\cite{CPH18}.
To fit the LECs of this new interaction, we consider the experimental energies of the five lowest states in the $^{11}$Be spectrum, and results of the \emph{ab initio} calculation of Calci \etal \cite{CNR16}.
For the bound states, we use their predictions of the ANC.
For the resonant states, we use the experimental widths.
To constrain the additional LEC in the $s$ and $p$ waves, we use the \emph{ab initio} prediction for the phase shift in these partial waves at low energy.

This natural extension of the work initiated in Ref.~\cite{CPH18} confirms the convergence of the Halo-EFT scheme at NLO for Coulomb breakup.
The N$^2$LO expansion also enables us to eliminate the cutoff dependency in the $p_{3/2}$ phase shift and hence to confirm that breakup cross sections are not sensitive to the short-range physics of the projectile.
Moreover, the excellent agreement with experiment validates the predictions of Calci \etal \cite{CNR16}.

This paper is structured as follows.
After a short presentation of the effective-potential theory at N$^2$LO (see \Sec{eft}), we detail in \Sec{Be11} how the LECs of that expansion have been fitted.
Section~\ref{bu} presents the results of our breakup calculations first on Pb at 69~MeV/nucleon (\Sec{PbE69}) and then on C at 67~MeV/nucleon (\Sec{CE67}).
We conclude this work in \Sec{conclusion}.

\section{{Effective-potential theory} at N$^2$LO}\label{eft}
To describe the breakup of $^{11}$Be, we consider the usual few-body model of the reaction.
The projectile $P$ is described as a halo neutron $n$ loosely bound to a $^{10}$Be core $c$ assumed in its $0^+$ ground state, and of which the internal structure is ignored.
The Hamiltonian for this two-body structure reads
\beq
H_0=-\frac{\hbar^2}{2\mu}\Delta+V(r),
\eeqn{e2}
where $\ve{r}$ is the $c$-$n$ relative coordinate and $\mu$ is the $c$-$n$ reduced mass.
The potential $V$ simulates the $^{10}$Be-$n$ interaction.
Following the original idea of Ref.~\cite{CPH18}, we consider an effective interaction inspired from Halo-EFT \cite{BHvK02,BHvK03}, see Ref.~\cite{HJP17} for a recent review.
However, as mentioned before, in the present work, we go one step further and consider a description of $^{11}$Be up to N$^2$LO. Because our reaction code requires $V$ to be local, we cannot consider the interaction as exactly derived from the Halo-EFT Lagrangian at N$^2$LO.
In particular, the sub-leading external currents have to be overlooked.
In this study, we rather use the effective-potential approach of Lepage \cite{Lepage97}.
This interaction is fitted partial wave per partial wave up to $d$ waves.

Expanding the development of Ref.~\cite{CPH18} following Lepage \cite{Lepage97}, we add a third term in the $^{10}$Be-$n$ effective interaction in the $s$ and $p$ waves
\beq
V_{lj}(r)&=& V_{lj}^{(0)}e^{-\frac{r^2}{2\sigma^2}}+V_{lj}^{(2)}\,r^2e^{-\frac{r^2}{2\sigma^2}}+V_{lj}^{(4)}\,r^4e^{-\frac{r^2}{2\sigma^2}},
\eeqn{e1}
where $l$ and $j$ are the quantum numbers of the orbital and total angular momenta, respectively.
The low-energy constants (LECs) $V_{lj}^{(0)}$, $V_{lj}^{(2)}$,  and $V_{lj}^{(4)}$ are fitted to know experimental observables, such as the neutron separation energy, or theoretical predictions.
The fitting procedure is detailed in \Sec{Be11}.

In the $d$ waves, however, we consider only the first two terms of the expression \eq{e1}; viz. $V_{dj}^{(4)}=0, \forall j$.
For $l>2$, the $c$-$n$ interaction is assumed to be nil.

In the usual few-body model of breakup, the target $T$ is seen as a structureless cluster.
Its interactions with the projectile constituents, $c$ and $n$, are described by optical potentials $V_{cT}$ and $V_{nT}$ chosen in the literature.
We consider the same interactions as in Ref.~\cite{CPH18}, see Sec.~VI\,A of that reference.

This model leads to the resolution of a three-body \Sch equation, of which the Hamiltonian reads
\beq
H=-\frac{\hbar^2}{2\mu_{PT}}\Delta_R+H_0+V_{cT}(R_{cT})+V_{nT}(R_{nT}),
\eeqn{e3}
where $\ve{R}$ is the relative coordinate between the projectile center of mass and the target, $\mu_{PT}$ is the $P$-$T$ reduced mass, and $R_{cT}$ and $R_{nT}$ are, respectively, the $c$-$T$ and $n$-$T$ distances.

Equation \eq{e3} has to be solved with the condition that the $^{11}$Be projectile is initially in its ground state $\phi_{n_0l_0j_0m_0}$ of energy $E_{n_0j_0l_0}$, with $n_0$ the number of nodes in the radial wave function.
Accordingly, the three-body wave function $\Psi$ behaves as
\beq
\Psi^{(m_0)}(\ve{r},\ve{R})\flim{Z}{-\infty}e^{iKZ+\ldots}\phi_{n_0l_0j_0m_0}(\ve{r}),
\eeqn{e4}
with the $Z$ coordinate chosen along the incoming beam axis, $m_0$ is the projection of the total angular momentum $j_0$ of the initial state, and the incoming wave number is related to the total energy $\cal E$ in the $P$-$T$ center-of-mass rest frame $\hbar^2K^2/2\mu_{PT}={\cal E}-E_{n_0l_0j_0}$.
The ``$\ldots$'' in the incoming wave indicates that, even at large distances, the $P$-$T$ motion is affected by the Coulomb interaction.

To solve \Eq{e3} with the initial condition \eq{e4}, we follow Ref.~\cite{CPH18} and use the dynamical eikonal approximation (DEA) \cite{BCG05,GBC06}.
This approximation is very successful in describing breakup reactions at intermediate energy.
In particular it is perfectly suited to study the Coulomb and nuclear-dominated breakup reactions measured at RIKEN \cite{Fuk04}, which we re-analyze in this paper.

\begin{table*}
\begin{tabular}{l|ccccc}\hline\hline
$\sigma$ & $V^{(0)}_{s1/2}$ & $V^{(2)}_{s1/2}$ & $V^{(4)}_{s1/2}$ & $E_{1s1/2}$ & ${\cal C}_{1s1/2}$ \\
(fm) & (MeV) & (MeV fm$^{-2}$) & (MeV fm$^{-4}$) & (MeV) & (fm$^{-1/2}$) \\ \hline
 1.2 & $-382.895$ & $151.3$ & $-13.0$ & $-0.5033$ & $0.7858$ \\
 1.5 & $-155.3167$ & $21.35$ & $-1.0$ & $-0.5031$ & $0.7858$ \\
 2    & $-100.7272$ & $7.84$ & $-0.15$ & $-0.5031$ & $0.7861$ \\ \hline
 \emph{Ab initio} &  &  &  & $-0.5$ & $0.786$ \\ \hline\hline
\end{tabular}
\caption{Depths of the effective $^{10}$Be-$n$ interaction \eq{e1} in the $s_{1/2}$ partial wave at N$^2$LO, and the corresponding ground-state energy and ANC. The corresponding \emph{ab initio} predictions of Calci \etal \cite{CNR16} used in the adjustment are listed in the last line.}\label{t1}
\end{table*}

\begin{figure*}[!ht]
        \centering
\includegraphics[width=8cm]{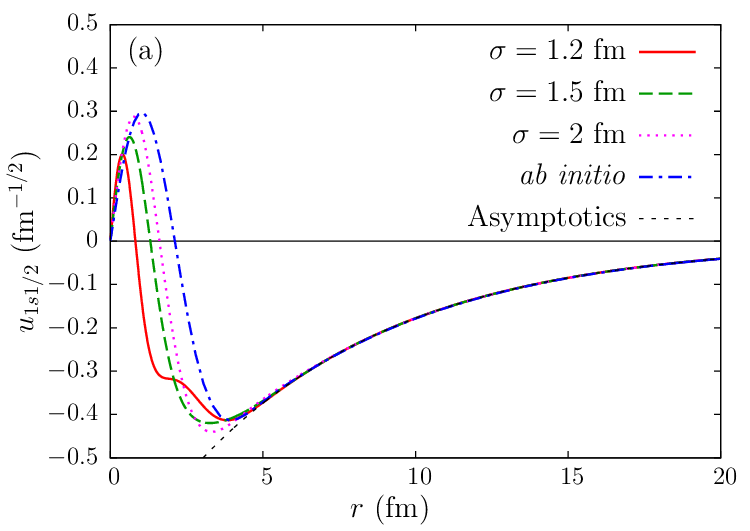}    
\includegraphics[width=8cm]{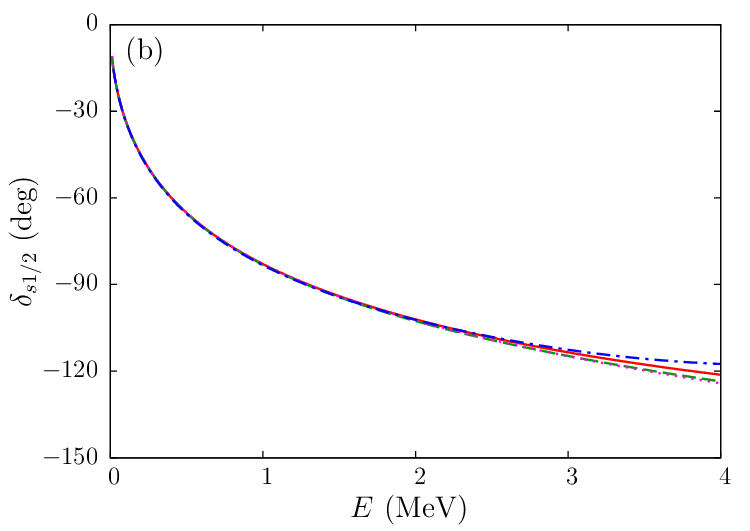}    
        \caption
        {Effective-potential calculations of $^{11}$Be in the $s_{1/2}$ partial wave at N$^2$LO.
        (a) Reduced radial wave function of the $\half^+$ ground state described as a $1s_{1/2}$ bound state in the potential given by \Eq{e1}.
        (b) Phase shift as a function of the $^{10}$Be-$n$ relative energy $E$.
        The results are shown for the cutoffs $\sigma=1.2$, 1.5, and 2~fm.
        The \emph{ab initio} predictions of Ref.~\cite{CNR16} against which the LECs of \Eq{e1} have been fitted are shown in blue dash-dotted lines.}\label{f1}
\end{figure*}

\section{Description of $^{11}$Be}\label{Be11}

\subsection{$s_{1/2}$ partial wave}\label{s1}

In the usual single-particle description of $^{11}$Be, the $\half^+$ ground state corresponds to a $1s_{1/2}$ neutron bound to a $^{10}$Be core in its $0^+$ ground state.
It is seen as an intruder from the $sd$ shell into the $p$ shell.

At N$^2$LO, the effective $^{10}$Be-$n$ potential includes three terms in the $s$ wave, see \Eq{e1}.
To fit the three LECs of that interaction, wee need three physical observables.
As in Ref.~\cite{CPH18}, we consider the experimental one-neutron separation energy $S_n(^{11}{\rm Be})=0.503$~MeV as the first constraint.
The second one is the ANC predicted by the \emph{ab initio} calculation of Calci \etal \cite{CNR16}: ${\cal C}_{1s1/2}=0.786$~fm$^{-1/2}$.
This value leads to excellent agreement with experiments on Coulomb and nuclear breakup of $^{11}$Be \cite{CPH18,MC19}, $^{10}{\rm Be}(d,p)$ transfer \cite{YC18}, and one-neutron knockout off $^{11}$Be \cite{HC21}.
It can therefore be considered as very reliable.
As third fitting observable, we consider the $\half^+$ phase shifts predicted by Calci \etal \cite{CNR16}.
At NLO, fixing  $S_n(^{11}{\rm Be})$ and ${\cal C}_{1s1/2}$ naturally produces an $s_{1/2}$ phase shift in agreement with the \emph{ab initio} calculation up to 1.5~MeV \cite{CPH18,SCB10}.
To extend the energy range of that agreement, we use the method of least squares to constrain the LECs and minimize the square of the residuals between the $s_{1/2}$ phase shifts of the effective potential \eq{e1} and the \emph{ab initio} ones up to $E=2$~MeV.

We adjust the LECs of this interaction considering the same three cutoffs as in Ref.~\cite{CPH18}: $\sigma=1.2$, 1.5, and 2~fm.
The results of this fitting are provided in \tbl{t1}, which lists, in addition to the depths $V_{s1/2}^{(0)}$, $V_{s1/2}^{(2)}$, and $V_{s1/2}^{(4)}$, the precise values obtained by these potentials for $S_n$ and ${\cal C}_{1s1/2}$.
Figure~\ref{f1} shows (a) the reduced radial wave function $u_{1s1/2}$ obtained from these fits, and (b) the corresponding phase shift $\delta_{s1/2}$.
The added value of the N$^2$LO fit can be observed by comparing these plots with, respectively, Figs.~1 and 2 of Ref.~\cite{CPH18}; for clarity, we use the same line style and color code as in that reference.
The calculations with $\sigma=1.2$ (solid red lines), 1.5 (green dashed lines), and 2~fm (magenta dotted lines) are compared to the \emph{ab initio} results (blue dash-dotted lines).

\begin{table*}[t]
\begin{tabular}{l|ccccc}\hline\hline
$\sigma$ & $V^{(0)}_{p1/2}$ & $V^{(2)}_{p1/2}$ & $V^{(4)}_{p1/2}$ & $E_{0p1/2}$ & ${\cal C}_{0p1/2}$ \\
(fm) & (MeV) & (MeV fm$^{-2}$) & (MeV fm$^{-4}$) & (MeV) & (fm${-1/2}$) \\ \hline
 1.2 & $361.7838$ & $-198.65$ & $15.4$ & $-0.1843$ & $0.1291$ \\
 1.5 & $-46.1204$ & $-9.7$ & $0.9$ & $-0.1841$ & $0.1290$ \\
 2    & $-64.0397$ & $5.065$ & $-0.065$ & $-0.1841$ & $0.1290$ \\ \hline
 \emph{Ab initio} &  &  &  & $-0.1848$ & $0.1291$ \\ \hline\hline
\end{tabular}
\caption{Depths of the effective $^{10}$Be-$n$ interaction \eq{e1} in the $p_{1/2}$ partial wave at N$^2$LO, and the corresponding energy and ANC. The corresponding \emph{ab initio} predictions of Calci \etal \cite{CNR16} used in the adjustment are listed in the last line.}\label{t2}
\end{table*}

 \begin{figure*}[!ht]
        \centering
\includegraphics[width=8cm]{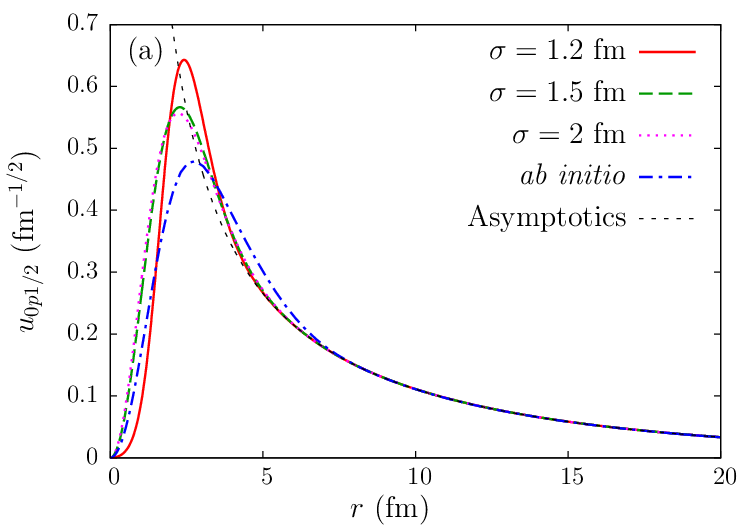}    
\includegraphics[width=8cm]{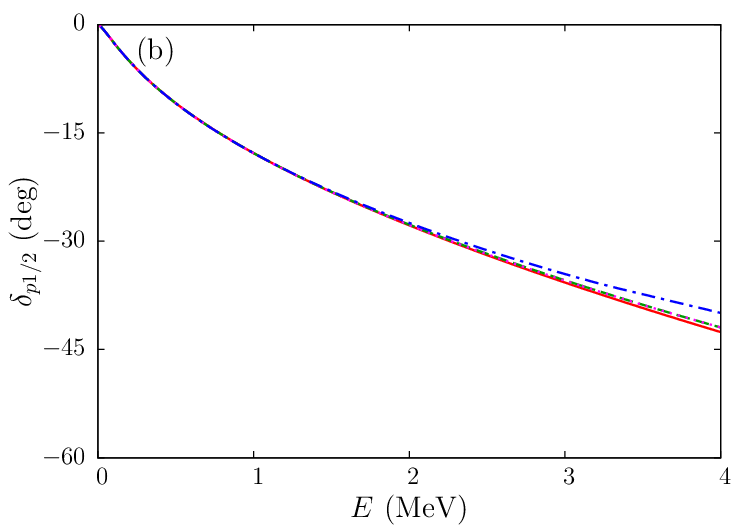}    
        \caption
        {Effective-potential calculations of $^{11}$Be in the $p_{1/2}$ partial wave at N$^2$LO.
        (a) Reduced radial wave function of the $\half^-$ excited state described as a $0p_{1/2}$ bound state in the potential given by \Eq{e1}.
        (b) Phase shift as a function of the $^{10}$Be-$n$ relative energy $E$.
        The results are shown for the cutoffs $\sigma=1.2$, 1.5, and 2~fm.
        The \emph{ab initio} predictions of Ref.~\cite{CNR16} against which the LECs of \Eq{e1} have been fitted are shown in blue dash-dotted lines.}\label{f2}
\end{figure*}

Thanks to the additional parameter in the N$^2$LO effective interaction, we extend the agreement with the \emph{ab initio} phase shift up to 2.5~MeV; so beyond the 0--2~MeV range within which it was fitted.
Interestingly, this improved description in the phase shift does not lead to wave functions much closer to the \emph{ab initio} overlap wave function.
However, the new radial wave functions leave their asymptotic behavior (thin dashed black line) at the same radius as the \emph{ab initio} one, around $r\simeq5$~fm.
At NLO, they held their asymptotic behavior down to smaller radii.
Interestingly, compared to the NLO interactions of Ref.~\cite{CPH18}, the N$^2$LO potentials lead to more significant differences between the different cutoffs in the interior of the wave function.
In particular the position of the node changes significantly with $\sigma$.
Accordingly, any small sensitivity to the internal part of the wave function will be noticeable in the breakup cross sections.

\subsection{$p_{1/2}$ partial wave}\label{p1}

Within an extreme shell model, the $p_{1/2}$ partial wave should host $^{11}$Be valence neutron in the ground state.
As mentioned earlier, it does not.
Instead, it corresponds to the $\half^-$ excited bound state of the nucleus.
Here also, the three LECs of the effective interaction are fitted to reproduce the experimental energy of that state, 184~keV under the one-neutron separation threshold, the \emph{ab initio} prediction for this state's ANC ${\cal C}_{0p1/2}=0.1291$~fm$^{-1/2}$, and the low-energy phase shift in that partial wave up to $E=2$~MeV \cite{CNR16}.
The fitted depths and precise values of the parameters they provide are listed in \tbl{t2}.
The reduced radial wave functions $u_{0p1/2}$ and phase shifts $\delta_{p1/2}$ are shown in Figs.~\ref{f2}(a) and (b), respectively.

In this partial wave, the changes compared to the NLO calculation of Ref.~\cite{CPH18} are striking.
Even though all NLO potentials reproduce reasonably well the \emph{ab initio} $p_{1/2}$ phase shift (see Fig.~4 of Ref.~\cite{CPH18}), the dependence of the corresponding $\delta_{p1/2}$ on the regulator $\sigma$ is significantly larger than in the $s_{1/2}$ partial wave.
Adding an extra term to the effective potential enables us to reduce that dependence to a level similar to the N$^2$LO $s_{1/2}$ potential [see \Fig{f1}(b)].
Here, all $\sigma$ provides nearly identical phase shift up to 3~MeV, and the agreement with the \emph{ab initio} prediction is excellent up to 2~MeV.

We observe that the additional term in the N$^2$LO potential reduces the difference between the radial wave functions down to $r\simeq4$~fm, see \Fig{f2}(a).
However, this does not improve the agreement between the effective single-particle wave function and the \emph{ab initio} overlap wave function in the range $r=4$--7~fm.
Clearly, even at N$^2$LO, the short-range nature of the effective interaction \eq{e1} cannot reproduce this feature of the \emph{ab initio} prediction.
This confirms the mean-field nature of this state \cite{PXS06}.
 
\subsection{$p_{3/2}$ partial wave}\label{p3}

\begin{table*}
\begin{tabular}{l|ccccc}\hline\hline
$\sigma$ & $V^{(0)}_{p3/2}$ & $V^{(2)}_{p3/2}$ & $V^{(4)}_{p3/2}$ & $E_{p3/2}$ & $\Gamma_{p3/2}$ \\
(fm) & (MeV) & (MeV fm$^{-2}$) & (MeV fm$^{-4}$) & (MeV) & (keV) \\ \hline
 1.2 & $-686.666$ & $210.06$ & $-7.3$ & $2.150$ & $212$ \\
 1.5 & $-498.698$ & $120.10$ & $-4.125$ & $2.150$ & $220$ \\
 2    & $-299.57$ & $45.00$ & $-1.0$ & $2.150$ & $217$ \\ \hline
 \emph{Ab initio} &  &  &  & $2.15$ & $190$ \\ 
 Experiment &  &  &  & $2.15\pm0.01$ & $206\pm8$ \\ \hline\hline
\end{tabular}
\caption{Depths of the effective $^{10}$Be-$n$ interaction \eq{e1} in the $p_{3/2}$ partial wave at N$^2$LO, and the corresponding energy and  width of the $\thal^-$ state. The \emph{ab initio} predictions of Calci \etal \cite{CNR16} and experimental values \cite{KKP12} used in the adjustment are listed in the last two lines.}\label{t3}
\end{table*}

The major source of uncertainty in the ``beyond-NLO'' reaction calculations presented in Sec.~VII of Ref.~\cite{CPH18} comes from the $p_{3/2}$ partial wave. In that article, the two LECs of the potential had been adjusted to reproduce the experimental energy and width of the $\thal^-$ resonance of $^{11}$Be.
However, with only two parameters, the resulting $p_{3/2}$ phase shift strongly depends on the cutoff $\sigma$ and differs significantly from the \emph{ab initio} prediction, see Fig.~10 of Ref.~\cite{CPH18}.
This dependency affects the Coulomb breakup cross section by more than 10\%, see Fig.~12(a) of Ref.~\cite{CPH18}.
To reproduce the \emph{ab initio} $\delta_{p3/2}$, the cutoff was used as an additional fitting parameter.
A good agreement was found using $\sigma=1$~fm.
This led to a good agreement with breakup data on both Pb and C, see Figs.~12(b) and 14.

\begin{figure}[!ht]
        \centering
\includegraphics[width=8cm]{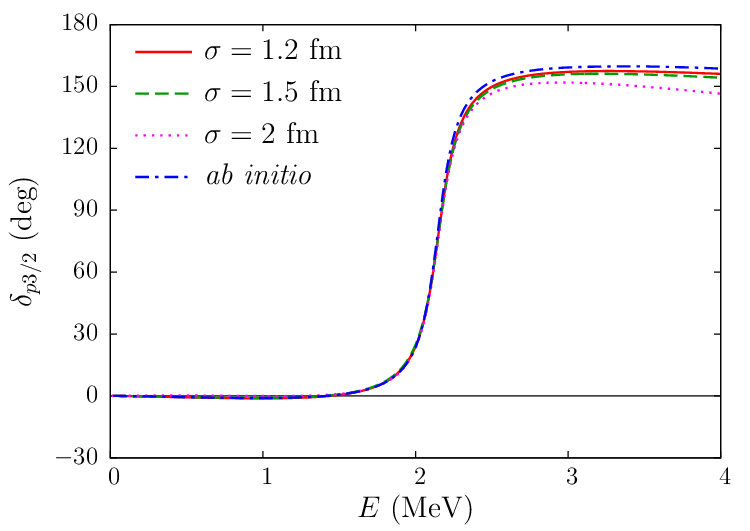}
        \caption
        {Effective-potential phase shift the $p_{3/2}$ partial wave at N$^2$LO as a function of the $^{10}$Be-$n$ relative energy $E$.
        The results are shown for the cutoffs $\sigma=1.2$, 1.5, and 2~fm.
        The \emph{ab initio} predictions of Ref.~\cite{CNR16} against which the LECs of \Eq{e1} have been fitted are shown in blue dash-dotted lines.}\label{f3}
\end{figure}

Such a dependence of the calculations on the cutoff is not expected in effective theories because the description of the nucleus should not depend on the short-range physics. One way to counter this is to extend the description to N$^2$LO.
We thus repeat the fit made in Ref.~\cite{CPH18}, but using the additional term in \Eq{e1} within a least-square method to reproduce the \emph{ab initio} $p_{3/2}$ phase shift at low-energy, viz. to force our $\delta_{p3/2}$ to stay close to zero in the range 0--1.5 MeV.
The corresponding LECs, as well as the position and width of the $p_{3/2}$ resonance, which they produce are listed in \tbl{t3}.
The $p_{3/2}$ phase shift is plotted in \Fig{f3} as a function of the $c$-$n$ relative energy $E$.

Adding an order to the effective expansion clearly solves the $\sigma$ dependency in the $p_{3/2}$ partial wave observed in Ref.~\cite{CPH18}.
The N$^2$LO interactions reproduce not only the energy and width of the known $\thal^-$ resonance, but the $p_{3/2}$ phase shift they generate is also independent of $\sigma$ up to nearly 2.5~MeV.
By construction they perfectly agree with the \emph{ab initio} prediction below that energy.

The $^{10}$Be-$n$ interactions constructed here within an effective-potential approach at N$^2$LO solve both issues mentioned in the Introduction.
First, they enable us to check that the effective expansion converges.
Second, they remove the significant dependence on the regulator $\sigma$ noted in Ref.~\cite{CPH18}.
In addition, they enable us to better reproduce the \emph{ab initio} predictions.
In particular, the phase shifts in the $s$ and $p$ waves are now in perfect agreement with the calculations of Calci \etal up to 2--3~MeV.
Accordingly, these new $^{10}$Be-$n$ interactions offer us the possibility to check that the effective scheme actually converges and to extend the test of the \emph{ab initio} prediction to larger energies in the $^{10}$Be-$n$ continuum.
In the next section, we test this on actual reaction data using accurate calculations of the Coulomb and nuclear breakup of $^{11}$Be.

\section{Breakup calculations of $^{11}$Be }\label{bu}
\subsection{On Pb at 69~MeV/nucleon}\label{PbE69}

Using the N$^2$LO $^{10}$Be-$n$ interactions developed in the previous section, we repeat the reaction calculation for the Coulomb breakup of $^{11}$Be on Pb at 69~MeV/nucleon.
Besides the $^{10}$Be-$n$  interaction, the inputs (optical potentials) and numerical conditions are identical to those of Ref.~\cite{CPH18}.
That reaction has been measured at RIKEN by Fukuda \etal \cite{Fuk04}.
We can thus compare our calculations to these accurate data to test the reliability of this extended effective description of $^{11}$Be.

\begin{figure}[!ht]
        \centering
\includegraphics[width=8cm]{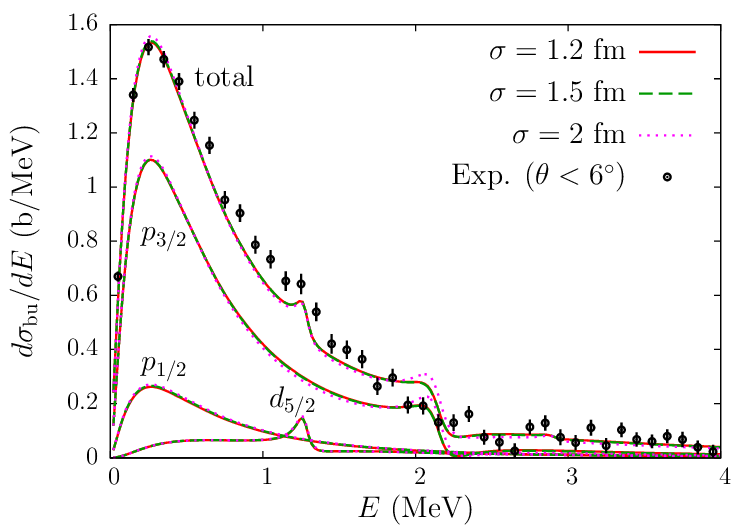}
        \caption
        {Theoretical breakup cross section of ${11}$Be on Pb at 69~MeV/nucleon plotted as a function of the $^{10}$Be-$n$ relative energy $E$ after dissociation.
        The contributions of the dominant partial waves $p_{3/2}$, $p_{1/2}$, and $d_{5/2}$ are shown as well.
        The calculations with the three cutoffs $\sigma=1.2$, 1.5, and 2~fm are nearly identical, illustrating the independence of the reaction process to the short-range physics of $^{11}$Be.
        The data of Ref.~\cite{Fuk04} are shown for comparison.}\label{f4}
\end{figure}

\begin{figure}[!ht]
        \centering
\includegraphics[width=8cm]{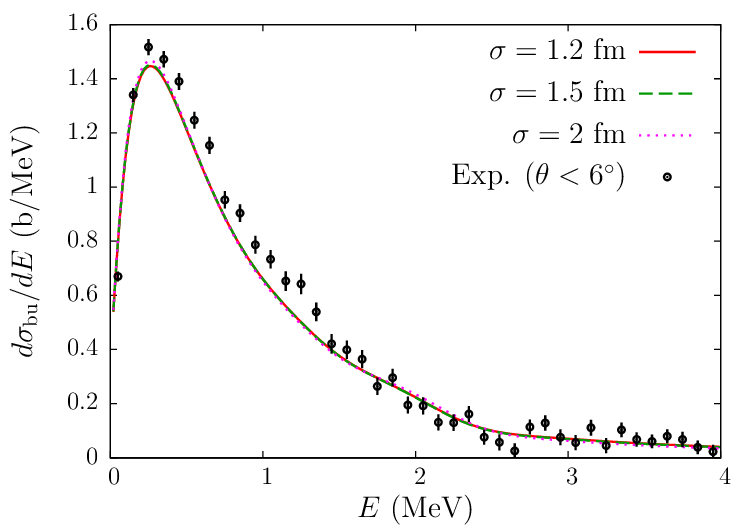}
        \caption
        {Breakup of $^{11}$Be on Pb at 69~MeV/nucleon.
        The theoretical cross sections of \Fig{f4} are compared to the data of Ref.~\cite{Fuk04} after folding with the experimental energy resolution.}\label{f5}
\end{figure}

Figure~\ref{f4} presents the breakup cross section as a function of the relative energy $E$ between the $^{10}$Be core and the neutron after dissociation.
It is equivalent to Fig.~12(a) of Ref.~\cite{CPH18}.
The DEA calculations are shown for the three regulators $\sigma=1.2$, 1.5, and 2~fm.
In addition to the total cross section, the dominant contributions of the $p_{3/2}$, $p_{1/2}$, and $d_{5/2}$ partial waves are shown as well.
To analyze these results without any filter, the theoretical results have not been folded with the experimental resolution; this is done in \Fig{f5}.
We nevertheless show the RIKEN data for comparison.

Compared to the calculations shown in Fig.~12(a) of Ref.~\cite{CPH18}, these N$^2$LO results exhibit nearly no sensitivity to the regulator $\sigma$.
This is true in all major partial-wave contributions.
The only slight sensitivity to $\sigma$ is noted at the energy of the $p_{3/2}$ resonance, where the breakup cross section obtained with $\sigma=2$~fm is a bit larger than the others.
This can be traced back to the small difference in the $p_{3/2}$ phase shift above 2~MeV, see \Fig{f3}.

This clear independence on $\sigma$ is obtained despite the strong variation observed in the internal part of the radial wave function of the $1s_{1/2}$ ground state, see \Fig{f1}(a).
This confirms previous results, which have shown the peripherality of breakup reactions, i.e. that they probe only the asymptotic part of the projectile wave functions \cite{CN07,CPH18}.
Accordingly they are sensitive to the ANC of the bound state and the phase shift in the continuum \cite{CN06}.
The good agreement with experiment confirms the validity of the \emph{ab initio} predictions of Calci \etal \cite{CNR16}. 
However, for a meaningful comparison with data, the theoretical cross section should be folded with the experimental resolution; this is performed in \Fig{f5}.

After folding, the small bumps due to the presence of resonances in the $d_{5/2}$ and $p_{3/2}$ partial waves are washed out.
The different calculations are barely distinguishable now, and agree very well with experiment.
We can thus conclude that extending the effective-potential description of $^{11}$Be to N$^2$LO enables us to efficiently remove the dependence of the reaction observables to the cutoff $\sigma$.
Moreover it shows that the expansion scheme has converged.
The breakup cross section obtained at N$^2$LO is not significantly better than the one obtained at NLO, see Fig.~6 of Ref.~\cite{CPH18}.
As stated in that previous work, a NLO description of $^{11}$Be is both necessary and sufficient to describe the Coulomb breakup of $^{11}$Be.
The presence of the $d$ and $p_{3/2}$ resonances in the nucleus' spectrum has no visible effect on the calculation.
Only a much precise energy resolution could reveal these features in the cross section.

This new analysis also confirms the reliability of the \emph{ab initio} predictions of Calci \etal \cite{CNR16}.
Knowing that this reaction is purely peripheral \cite{CN06,CN07,CPH18}, and that the structure observables, which affect the calculations, correspond to these predictions, the excellent agreement we obtain with the data validates the predictions of Ref.~\cite{CNR16}.

\subsection{On C at 67~MeV/nucleon}\label{CE67}

We extend our analysis to nuclear-dominated breakup.
That reaction is also peripheral \cite{CN07}, but it is more sensitive to the resonant continuum \cite{Fuk04,CGB04}.
In Ref.~\cite{CPH18}, in an attempt to describe the RIKEN data \cite{Fuk04}, the $\fial^+$ and $\thal^+$ resonances have been added as single-particle resonances within the $d_{5/2}$ and $d_{3/2}$ partial waves, respectively.
Following the natural power counting of Halo-EFT, this description of $^{11}$Be goes ``beyond NLO''.
In the present work, we use the effective-potential description of the projectile at N$^2$LO, which includes these two $d$-wave resonances.
Our goal is to confirm the convergence of the Halo-EFT scheme and check that, at N$^2$LO, the dissociation on a light target are also independent of the cutoff.

\begin{figure}[!ht]
        \centering
\includegraphics[width=8cm]{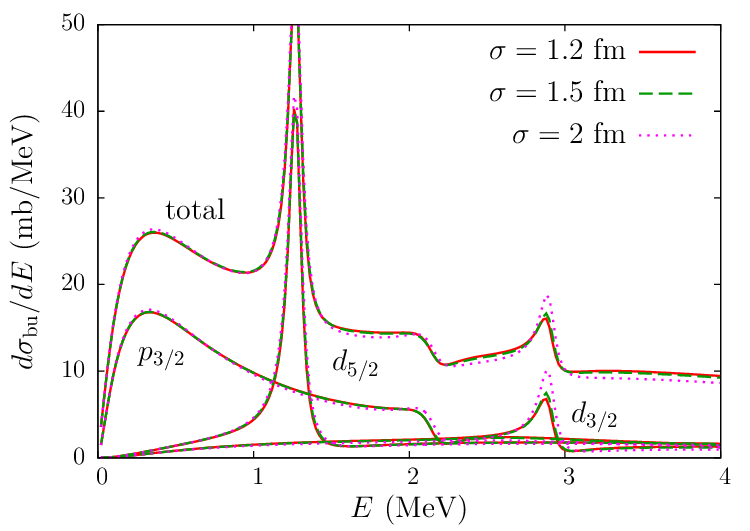}
        \caption
        {Theoretical breakup cross section of $^{11}$Be on C at 67~MeV/nucleon plotted as a function of the $^{10}$Be-$n$ relative energy $E$ after dissociation.
        The contributions of the dominant partial waves $p_{3/2}$, $d_{5/2}$, and $d_{3/2}$ are shown as well.
        The near-equality of the results obtained with the three different cutoffs $\sigma=1.2$, 1.5, and 2~fm shows the independence of the reaction calculation to the short-range physics of $^{11}$Be.}\label{f6}
\end{figure}

Figure \ref{f6} displays the results of our DEA calculations for the collision of $^{11}$Be on C at 67~MeV/nucleon, which corresponds to the conditions of the RIKEN experiment \cite{Fuk04}.
For the three cutoffs, the breakup cross section is plotted as a function of the relative energy $E$ between the $^{10}$Be core and the halo neutron $n$ after dissociation.
In addition to the total cross section, the main partial-wave contributions ($p_{3/2}$, $d_{5/2}$, and $d_{3/2}$) are shown separately.

As already discussed in Refs.~\cite{Fuk04,CGB04,CPH18}, the presence of the $d_{5/2}$ resonance significantly affects the breakup process.
It favors the dissociation of $^{11}$Be at that energy leading to a clear and significant peak at that resonant state.
A similar effect, though much less ample is also observed around the $d_{3/2}$ resonance. Using the N$^2$LO description developed in \Sec{Be11}, the breakup cross section is completely independent of $\sigma$, but in the $d$-wave resonances as already noted in Refs.~\cite{CN07,CPH18}.
This result confirms the convergence of the effective-potential scheme; going to N$^2$LO does not lead to significant change compared to the previous calculation, see Fig.~13 of Ref.~\cite{CPH18}.
It also shows that, although being dominated by the short-ranged nuclear interaction, this reaction is also purely peripheral \cite{CN07} and, as such, does not depend on the short-range physics of the projectile, but for the resonant breakup \cite{CPH18}.

\begin{figure}[!ht]
        \centering
\includegraphics[width=8cm]{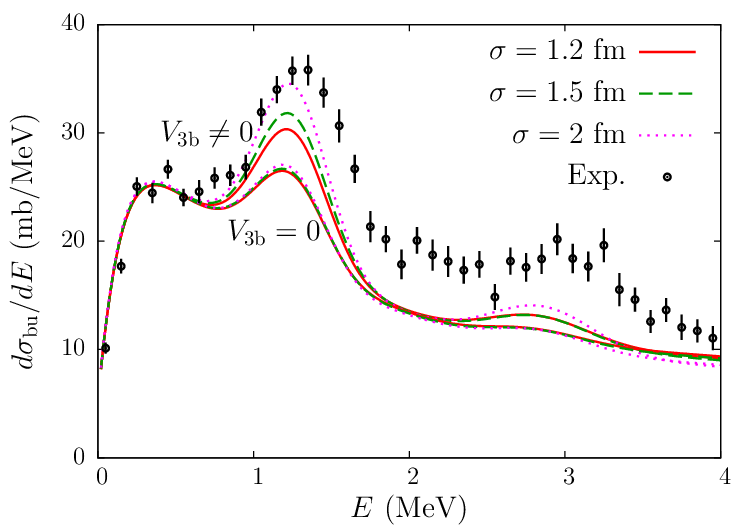}
        \caption
        {Nuclear breakup of $^{11}$Be on C at 67~MeV/nucleon; comparison of the theoretical cross sections with the experimental data of Ref.~\cite{Fuk04}. Our calculations are shown without (V$_{\text{3b}}=0$, lower curves) and with (V$_{\text{3b}}\ne 0$, upper curves) the effective three-body force simulating core excitation \cite{CPH22}. The theoretical cross sections have been folded with the experimental energy resolution \cite{Fuk04}.}\label{f7}
\end{figure}

To compare our calculations with the experiment, we fold our theoretical results with the energy resolution provided in Ref.~\cite{Fuk04}.
The corresponding cross sections are shown in \Fig{f7} alongside the data; see the bottom curves labelled $V_{\rm 3b}=0$.
They reproduce the general features of the experimental cross sections, but miss the peaks at the $\fial^+$ and $\thal^+$ energies. The folding has smoothened the narrow peaks of the $d$-wave resonances seen in \Fig{f6}.
The $d_{3/2}$ one is completely washed out, and the $d_{5/2}$ one is not high enough to correctly reproduce the data.

As shown by Moro and Lay, a significant part of the resonant breakup comes from the excitation of the $^{10}$Be core \cite{ML12}.
The role played by the $2^+_1$ excited state of $^{10}$Be dominates the breakup towards the $\thal^+$ resonance.
The role played by the core excitation in the breakup towards the $\fial^+$ is smaller, yet non negligible.
This degree of freedom is not included within our effective theory, which explains why we cannot reproduce the data.

The virtual excitation of the core during the breakup reaction can be simulated by an  effective three-body force \cite{CPH22}
\beq
\lefteqn{V_{3b}(\ve{R}_{cT},\ve{r}) = V_0^{3b}(R_0^{cT},R_0^{cn}) \, Y_2^0\left(\widehat R_{cT}\cdot\widehat r,0\right)}\nonumber\\
&\times& \left(\frac{R_{cT}}{R_0^{cT}}\right)^2 e^{-\left(\frac{R_{cT}}{R_0^{cT}}\right)^2} \left(\frac{r}{R_0^{cn}}\right)^2 e^{-\left(\frac{r}{R_0^{cn}}\right)^2}.
\eeqn{e3b}
We therefore repeat the calculations using that three-body force with the ranges $R_0^{cn}=\sqrt{2}\sigma$, $R_0^{cT}=3.5$~fm, and a depth $V_{\rm 3b}=-100$~MeV \cite{CPH22}.
This significantly increases the breakup strength in the energy ranges of both resonances; see the upper curves labelled $V_{\rm 3b}\ne0$ in \Fig{f7}.
In particular, the cross section obtained for the regulator $\sigma=2$~fm is in fair agreement with the data at the energy of the $\fial^+$ resonance; the other two cutoffs lead to too small resonant-breakup cross sections.
However, the results of Ref.~\cite{CPH22} suggest that this depth can be adjusted to reproduce the data.
The $\sigma$-dependence of these new results confirms that the resonant breakup is dominated by short-ranged physics, which is not included in the effective theory considered here.
Including explicitly the core excitation in the model of $^{11}$Be is the best way to confirm that \cite{ML12}.
As shown by the present and previous \cite{CPH18,YC18,MC19,MYC19,HC21} studies, effective approaches are useful tools to analyze reactions involving halo nuclei.
Extending this framework to add the core degree of freedom would certainly be an asset in the description of nuclear reactions.
A first step has already been taken in this direction \cite{KC24}.

\section{Conclusion}\label{conclusion}

Effective theories are very powerful to analyze nuclear reactions involving halo nuclei \cite{CPH18}.
Their very systematic expansion scheme offers a natural scaling of the structure observables, which influence the reaction cross sections.
This interesting feature has been observed for various reactions, from breakup \cite{CPH18,MC19} to transfer \cite{YC18} and knockout \cite{HC21}, and on different nuclei \cite{CPH18,MC19,YC18,MYC19,CP23,HC21}.
Moreover the LECs of the expansion can be fitted to predictions of accurate structure calculations.
Effective potentials therefore enable us to test the validity of these predictions by confronting them to experimental data.

The attractive features of effective descriptions of halo nuclei however requires that (i) the expansion schemes has properly converged \cite{Gri22}, and (ii) the result of the fit of the LECs be insensitive to the regulator $\sigma$.
In Ref.~\cite{CPH18}, we have obtained excellent agreement with reaction data for the Coulomb breakup of $^{11}$Be at 69~MeV/nucleon using a NLO description of the projectile.
The same description fails to reproduce the breakup cross section on $^{12}$C at 67~MeV/nucleon in the energy region of the $\fial^+$ and $\thal^+$ resonances.
The extension of that description ``beyond NLO'' proposed in Ref.~\cite{CPH18} to solve that issue leads to a significant sensitivity to the cutoff $\sigma$.
Such a seeming dependence on the short-range physics of the nucleus is not expected in effective theories.

To ensure the convergence of the effective expansion at NLO for Coulomb breakup and avoid the $\sigma$ dependency of our reaction calculations, we propose an effective-potential description of $^{11}$Be at N$^2$LO and run DEA calculations for the breakup of $^{11}$Be on both Pb and C.
Ideally, this study should have been performed within pure Halo-EFT.
However, since at present non-locality cannot be handled within our reaction code, we instead follow Lepage and use an effective-potential approach \cite{Lepage97}.
In the future, were Halo-EFT extended to N$^2$LO, efforts could be accomplished to include non-local interactions within our reaction code.

The addition of a term in the effective potential \eq{e1} enables us to reduce significantly the dependence of our structure observables to the regulator $\sigma$.
This is particularly true for the $p_{3/2}$ phase shift, for which $\sigma$ had to be used as a fitting parameter in Ref.~\cite{CPH18}.
Moreover, these new N$^2$LO potentials significantly extend the agreement between our single-particle description of $^{11}$Be and its \emph{ab initio} calculation by Calci \etal \cite{CNR16}.

The breakup cross sections obtained with this new effective description of $^{11}$Be are now nearly independent of the regulator $\sigma$.
Extending the description to N$^2$LO removes the seeming dependence to the short-range physics observed in Ref.~\cite{CPH18}.
Moreover, this result is obtained despite the significant differences observed in the initial bound-state wave functions for $r\lesssim 5$~fm.
This confirms, if that were still needed, that these reactions are purely peripheral, in the sense that they probe only the tail of the wave functions: the ANC in the ground state and the phase shift in the continuum.

On the lead target, our N$^2$LO calculations are nearly identical to the NLO calculations of Ref.~\cite{CPH18}, confirming that for this reaction a NLO description of $^{11}$Be is both needed, as shown in Ref.~\cite{CPH18}, and sufficient, as demonstrated by the present work.

The breakup reaction on the carbon target cannot be properly described at NLO.
The significant enhancement of the breakup strength at the resonance energies cannot be accounted for without including these states in the expansion scheme.
In a single-particle description, they correspond to one-neutron $d$ resonances, and therefore require an expansion up to, at least, N$^2$LO .
As already observed in Ref.~\cite{CPH18}, adding these single-particle resonances is not enough to fully reproduce the data.
Moro and Lay have shown that a significant amount of the breakup cross section can be related to the core excitation \cite{ML12}.
This effect can also be effectively simulated by a three-body force \cite{CPH22}.
Our present analysis shows that the strength of the three-body force depends on the cutoff $\sigma$ of the $^{10}$Be-$n$ interaction.
This is not surprising since this effect corresponds to physics within the projectile description.
Ideally, this resonant breakup should be studied using a description that explicitly accounts for that degree of freedom.
Such a description is currently under development \cite{KC24}.

In conclusion, we have demonstrated that to describe the Coulomb breakup of $^{11}$Be at intermediate energy, a NLO description of the projectile is both needed and sufficient.
For that reaction a N$^2$LO description does not improve the reaction calculation.
To properly describe the breakup on $^{12}$C, however, including the $d$-wave resonances is necessary.
This requires a N$^2$LO description.
This extended description also enables us to avoid the $\sigma$ dependency observed in $p_{3/2}$ partial wave at beyond-NLO in Ref.~\cite{CPH18}.
However, even at N$^2$LO the breakup strength at the $\fial^+$ and $\thal^+$ resonances is not enough to reproduce the RIKEN data.
As shown by Moro and Lay \cite{ML12}, this requires to explicitly account for the core excitation \cite{KC24}. 

\section*{Acknowledgement}
We thank H.\ W.\ Grie\ss hammer for his constructive comments on the previous NLO calculations.
This project has received funding from the Deutsche Forschungsgemeinschaft within the Collaborative Research Center SFB 1245 (Projektnummer 279384907) and the PRISMA+ (Precision Physics, Fundamental Interactions and Structure of Matter) Cluster of Excellence.
This research was supported in part by grant NSF PHY-1748958 to the Kavli Institute for Theoretical Physics (KITP).
\bibliography{N2LO}

\begin{thebibliography}{28}%
\makeatletter
\providecommand \@ifxundefined [1]{%
 \@ifx{#1\undefined}
}%
\providecommand \@ifnum [1]{%
 \ifnum #1\expandafter \@firstoftwo
 \else \expandafter \@secondoftwo
 \fi
}%
\providecommand \@ifx [1]{%
 \ifx #1\expandafter \@firstoftwo
 \else \expandafter \@secondoftwo
 \fi
}%
\providecommand \natexlab [1]{#1}%
\providecommand \enquote  [1]{``#1''}%
\providecommand \bibnamefont  [1]{#1}%
\providecommand \bibfnamefont [1]{#1}%
\providecommand \citenamefont [1]{#1}%
\providecommand \href@noop [0]{\@secondoftwo}%
\providecommand \href [0]{\begingroup \@sanitize@url \@href}%
\providecommand \@href[1]{\@@startlink{#1}\@@href}%
\providecommand \@@href[1]{\endgroup#1\@@endlink}%
\providecommand \@sanitize@url [0]{\catcode `\\12\catcode `\$12\catcode
  `\&12\catcode `\#12\catcode `\^12\catcode `\_12\catcode `\%12\relax}%
\providecommand \@@startlink[1]{}%
\providecommand \@@endlink[0]{}%
\providecommand \url  [0]{\begingroup\@sanitize@url \@url }%
\providecommand \@url [1]{\endgroup\@href {#1}{\urlprefix }}%
\providecommand \urlprefix  [0]{URL }%
\providecommand \Eprint [0]{\href }%
\providecommand \doibase [0]{http://dx.doi.org/}%
\providecommand \selectlanguage [0]{\@gobble}%
\providecommand \bibinfo  [0]{\@secondoftwo}%
\providecommand \bibfield  [0]{\@secondoftwo}%
\providecommand \translation [1]{[#1]}%
\providecommand \BibitemOpen [0]{}%
\providecommand \bibitemStop [0]{}%
\providecommand \bibitemNoStop [0]{.\EOS\space}%
\providecommand \EOS [0]{\spacefactor3000\relax}%
\providecommand \BibitemShut  [1]{\csname bibitem#1\endcsname}%
\let\auto@bib@innerbib\@empty
\bibitem [{\citenamefont {Tanihata}(1996)}]{Tan96}%
  \BibitemOpen
  \bibfield  {author} {\bibinfo {author} {\bibfnamefont {I.}~\bibnamefont
  {Tanihata}},\ }\href {\doibase 10.1088/0954-3899/22/2/004} {\bibfield
  {journal} {\bibinfo  {journal} {J. Phys. G}\ }\textbf {\bibinfo {volume}
  {22}},\ \bibinfo {pages} {157} (\bibinfo {year} {1996})}\BibitemShut
  {NoStop}%
\bibitem [{\citenamefont {Hansen}\ and\ \citenamefont {Jonson}(1987)}]{HJ87}%
  \BibitemOpen
  \bibfield  {author} {\bibinfo {author} {\bibfnamefont {P.~G.}\ \bibnamefont
  {Hansen}}\ and\ \bibinfo {author} {\bibfnamefont {B.}~\bibnamefont
  {Jonson}},\ }\href {\doibase 10.1209/0295-5075/4/4/005} {\bibfield  {journal}
  {\bibinfo  {journal} {Europhys. Lett.}\ }\textbf {\bibinfo {volume} {4}},\
  \bibinfo {pages} {409} (\bibinfo {year} {1987})}\BibitemShut {NoStop}%
\bibitem [{\citenamefont {Al-Khalili}\ and\ \citenamefont
  {Nunes}(2003)}]{AN03}%
  \BibitemOpen
  \bibfield  {author} {\bibinfo {author} {\bibfnamefont {J.}~\bibnamefont
  {Al-Khalili}}\ and\ \bibinfo {author} {\bibfnamefont {F.}~\bibnamefont
  {Nunes}},\ }\href {\doibase 10.1088/0954-3899/29/11/R01} {\bibfield
  {journal} {\bibinfo  {journal} {J. Phys. G}\ }\textbf {\bibinfo {volume}
  {29}},\ \bibinfo {pages} {R89} (\bibinfo {year} {2003})}\BibitemShut
  {NoStop}%
\bibitem [{\citenamefont {Baye}\ and\ \citenamefont {Capel}(2012)}]{BC12}%
  \BibitemOpen
  \bibfield  {author} {\bibinfo {author} {\bibfnamefont {D.}~\bibnamefont
  {Baye}}\ and\ \bibinfo {author} {\bibfnamefont {P.}~\bibnamefont {Capel}},\
  }\bibfield  {booktitle} {\emph {\bibinfo {booktitle} {Clusters in Nuclei,
  Vol. 2}},\ }\href {\doibase 10.1007/978-3-642-24707-1\_3} {\bibfield
  {journal} {\bibinfo  {journal} {Lecture Notes in Physics}\ }\textbf {\bibinfo
  {volume} {848}},\ \bibinfo {pages} {121} (\bibinfo {year} {2012})},\ \bibinfo
  {note} {{Ed. C. Beck}}\BibitemShut {NoStop}%
\bibitem [{\citenamefont {Bertulani}\ \emph {et~al.}(2002)\citenamefont
  {Bertulani}, \citenamefont {Hammer},\ and\ \citenamefont {{van
  Kolck}}}]{BHvK02}%
  \BibitemOpen
  \bibfield  {author} {\bibinfo {author} {\bibfnamefont {C.}~\bibnamefont
  {Bertulani}}, \bibinfo {author} {\bibfnamefont {H.-W.}\ \bibnamefont
  {Hammer}}, \ and\ \bibinfo {author} {\bibfnamefont {U.}~\bibnamefont {{van
  Kolck}}},\ }\href {\doibase https://doi.org/10.1016/S0375-9474(02)01270-8}
  {\bibfield  {journal} {\bibinfo  {journal} {Nucl. Phys.}\ }\textbf {\bibinfo
  {volume} {A712}},\ \bibinfo {pages} {37} (\bibinfo {year}
  {2002})}\BibitemShut {NoStop}%
\bibitem [{\citenamefont {Bedaque}\ \emph {et~al.}(2003)\citenamefont
  {Bedaque}, \citenamefont {Hammer},\ and\ \citenamefont {{van
  Kolck}}}]{BHvK03}%
  \BibitemOpen
  \bibfield  {author} {\bibinfo {author} {\bibfnamefont {P.}~\bibnamefont
  {Bedaque}}, \bibinfo {author} {\bibfnamefont {H.-W.}\ \bibnamefont {Hammer}},
  \ and\ \bibinfo {author} {\bibfnamefont {U.}~\bibnamefont {{van Kolck}}},\
  }\href {\doibase https://doi.org/10.1016/j.physletb.2003.07.049} {\bibfield
  {journal} {\bibinfo  {journal} {Phys. Lett.}\ }\textbf {\bibinfo {volume}
  {B569}},\ \bibinfo {pages} {159} (\bibinfo {year} {2003})}\BibitemShut
  {NoStop}%
\bibitem [{\citenamefont {Hammer}\ \emph {et~al.}(2017)\citenamefont {Hammer},
  \citenamefont {Ji},\ and\ \citenamefont {Phillips}}]{HJP17}%
  \BibitemOpen
  \bibfield  {author} {\bibinfo {author} {\bibfnamefont {H.-W.}\ \bibnamefont
  {Hammer}}, \bibinfo {author} {\bibfnamefont {C.}~\bibnamefont {Ji}}, \ and\
  \bibinfo {author} {\bibfnamefont {D.~R.}\ \bibnamefont {Phillips}},\ }\href
  {\doibase 10.1088/1361-6471/aa83db} {\bibfield  {journal} {\bibinfo
  {journal} {J. Phys. G}\ }\textbf {\bibinfo {volume} {44}},\ \bibinfo {pages}
  {103002} (\bibinfo {year} {2017})}\BibitemShut {NoStop}%
\bibitem [{\citenamefont {Capel}\ \emph {et~al.}(2018)\citenamefont {Capel},
  \citenamefont {Phillips},\ and\ \citenamefont {Hammer}}]{CPH18}%
  \BibitemOpen
  \bibfield  {author} {\bibinfo {author} {\bibfnamefont {P.}~\bibnamefont
  {Capel}}, \bibinfo {author} {\bibfnamefont {D.~R.}\ \bibnamefont {Phillips}},
  \ and\ \bibinfo {author} {\bibfnamefont {H.-W.}\ \bibnamefont {Hammer}},\
  }\href {\doibase 10.1103/PhysRevC.98.034610} {\bibfield  {journal} {\bibinfo
  {journal} {Phys. Rev. C}\ }\textbf {\bibinfo {volume} {98}},\ \bibinfo
  {pages} {034610} (\bibinfo {year} {2018})}\BibitemShut {NoStop}%
\bibitem [{\citenamefont {Yang}\ and\ \citenamefont {Capel}(2018)}]{YC18}%
  \BibitemOpen
  \bibfield  {author} {\bibinfo {author} {\bibfnamefont {J.}~\bibnamefont
  {Yang}}\ and\ \bibinfo {author} {\bibfnamefont {P.}~\bibnamefont {Capel}},\
  }\href {\doibase 10.1103/PhysRevC.98.054602} {\bibfield  {journal} {\bibinfo
  {journal} {Phys. Rev. C}\ }\textbf {\bibinfo {volume} {98}},\ \bibinfo
  {pages} {054602} (\bibinfo {year} {2018})}\BibitemShut {NoStop}%
\bibitem [{\citenamefont {Hebborn}\ and\ \citenamefont {Capel}(2021)}]{HC21}%
  \BibitemOpen
  \bibfield  {author} {\bibinfo {author} {\bibfnamefont {C.}~\bibnamefont
  {Hebborn}}\ and\ \bibinfo {author} {\bibfnamefont {P.}~\bibnamefont
  {Capel}},\ }\href {\doibase 10.1103/PhysRevC.104.024616} {\bibfield
  {journal} {\bibinfo  {journal} {Phys. Rev. C}\ }\textbf {\bibinfo {volume}
  {104}},\ \bibinfo {pages} {024616} (\bibinfo {year} {2021})}\BibitemShut
  {NoStop}%
\bibitem [{\citenamefont {Moschini}\ \emph {et~al.}(2019)\citenamefont
  {Moschini}, \citenamefont {Yang},\ and\ \citenamefont {Capel}}]{MYC19}%
  \BibitemOpen
  \bibfield  {author} {\bibinfo {author} {\bibfnamefont {L.}~\bibnamefont
  {Moschini}}, \bibinfo {author} {\bibfnamefont {J.}~\bibnamefont {Yang}}, \
  and\ \bibinfo {author} {\bibfnamefont {P.}~\bibnamefont {Capel}},\ }\href
  {\doibase 10.1103/PhysRevC.100.044615} {\bibfield  {journal} {\bibinfo
  {journal} {Phys. Rev. C}\ }\textbf {\bibinfo {volume} {100}},\ \bibinfo
  {pages} {044615} (\bibinfo {year} {2019})}\BibitemShut {NoStop}%
\bibitem [{\citenamefont {Capel}\ \emph {et~al.}(2023)\citenamefont {Capel}, ,
  \citenamefont {Phillips}, \citenamefont {Andis}, \citenamefont {Bagnarol},
  \citenamefont {Behzadmoghaddam}, \citenamefont {Bonaiti}, \citenamefont
  {Bubna}, \citenamefont {Capitani}, \citenamefont {Duerinck}, \citenamefont
  {Durant}, \citenamefont {Döpper}, \citenamefont {Boustani}, \citenamefont
  {Farrell}, \citenamefont {Geiger}, \citenamefont {Gennari}, \citenamefont
  {Goldberg}, \citenamefont {Herko}, \citenamefont {Kirchner}, \citenamefont
  {Kubushishi}, \citenamefont {Li}, \citenamefont {Muli}, \citenamefont {Long},
  \citenamefont {Martin}, \citenamefont {Mohseni}, \citenamefont {Moumene},
  \citenamefont {Paracone}, \citenamefont {Parnes}, \citenamefont {Romeo},
  \citenamefont {Springer}, \citenamefont {Svensson}, \citenamefont {Thim},\
  and\ \citenamefont {Yapa}}]{CP23}%
  \BibitemOpen
  \bibfield  {author} {\bibinfo {author} {\bibfnamefont {P.}~\bibnamefont
  {Capel}}, , \bibinfo {author} {\bibfnamefont {D.~R.}\ \bibnamefont
  {Phillips}}, \bibinfo {author} {\bibfnamefont {A.}~\bibnamefont {Andis}},
  \bibinfo {author} {\bibfnamefont {M.}~\bibnamefont {Bagnarol}}, \bibinfo
  {author} {\bibfnamefont {B.}~\bibnamefont {Behzadmoghaddam}}, \bibinfo
  {author} {\bibfnamefont {F.}~\bibnamefont {Bonaiti}}, \bibinfo {author}
  {\bibfnamefont {R.}~\bibnamefont {Bubna}}, \bibinfo {author} {\bibfnamefont
  {Y.}~\bibnamefont {Capitani}}, \bibinfo {author} {\bibfnamefont {P.-Y.}\
  \bibnamefont {Duerinck}}, \bibinfo {author} {\bibfnamefont {V.}~\bibnamefont
  {Durant}}, \bibinfo {author} {\bibfnamefont {N.}~\bibnamefont {Döpper}},
  \bibinfo {author} {\bibfnamefont {A.~E.}\ \bibnamefont {Boustani}}, \bibinfo
  {author} {\bibfnamefont {R.}~\bibnamefont {Farrell}}, \bibinfo {author}
  {\bibfnamefont {M.}~\bibnamefont {Geiger}}, \bibinfo {author} {\bibfnamefont
  {M.}~\bibnamefont {Gennari}}, \bibinfo {author} {\bibfnamefont
  {N.}~\bibnamefont {Goldberg}}, \bibinfo {author} {\bibfnamefont
  {J.}~\bibnamefont {Herko}}, \bibinfo {author} {\bibfnamefont
  {T.}~\bibnamefont {Kirchner}}, \bibinfo {author} {\bibfnamefont {L.-P.}\
  \bibnamefont {Kubushishi}}, \bibinfo {author} {\bibfnamefont
  {Z.}~\bibnamefont {Li}}, \bibinfo {author} {\bibfnamefont {S.~S.~L.}\
  \bibnamefont {Muli}}, \bibinfo {author} {\bibfnamefont {A.}~\bibnamefont
  {Long}}, \bibinfo {author} {\bibfnamefont {B.}~\bibnamefont {Martin}},
  \bibinfo {author} {\bibfnamefont {K.}~\bibnamefont {Mohseni}}, \bibinfo
  {author} {\bibfnamefont {I.}~\bibnamefont {Moumene}}, \bibinfo {author}
  {\bibfnamefont {N.}~\bibnamefont {Paracone}}, \bibinfo {author}
  {\bibfnamefont {E.}~\bibnamefont {Parnes}}, \bibinfo {author} {\bibfnamefont
  {B.}~\bibnamefont {Romeo}}, \bibinfo {author} {\bibfnamefont
  {V.}~\bibnamefont {Springer}}, \bibinfo {author} {\bibfnamefont
  {I.}~\bibnamefont {Svensson}}, \bibinfo {author} {\bibfnamefont
  {O.}~\bibnamefont {Thim}}, \ and\ \bibinfo {author} {\bibfnamefont
  {N.}~\bibnamefont {Yapa}},\ }\href {\doibase
  https://doi.org/10.1140/epja/s10050-023-01181-7} {\bibfield  {journal}
  {\bibinfo  {journal} {Eur. Phys. J. A}\ }\textbf {\bibinfo {volume} {59}},\
  \bibinfo {pages} {273} (\bibinfo {year} {2023})}\BibitemShut {NoStop}%
\bibitem [{\citenamefont {Capel}\ and\ \citenamefont {Nunes}(2007)}]{CN07}%
  \BibitemOpen
  \bibfield  {author} {\bibinfo {author} {\bibfnamefont {P.}~\bibnamefont
  {Capel}}\ and\ \bibinfo {author} {\bibfnamefont {F.~M.}\ \bibnamefont
  {Nunes}},\ }\href {\doibase 10.1103/PhysRevC.75.054609} {\bibfield  {journal}
  {\bibinfo  {journal} {Phys. Rev. C}\ }\textbf {\bibinfo {volume} {75}},\
  \bibinfo {pages} {054609} (\bibinfo {year} {2007})}\BibitemShut {NoStop}%
\bibitem [{\citenamefont {hammer}(2022)}]{Gri22}%
  \BibitemOpen
  \bibfield  {author} {\bibinfo {author} {\bibfnamefont {H.~W.~G.}\
  \bibnamefont {hammer}},\ }\href@noop {} {} (\bibinfo {year} {2022}),\
  \bibinfo {note} {comment during the program \emph{``Living Near Unitarity''}
  held at the KITP in Santa Barbara (California, USA)}\BibitemShut {NoStop}%
\bibitem [{\citenamefont {Lepage}(1997)}]{Lepage97}%
  \BibitemOpen
  \bibfield  {author} {\bibinfo {author} {\bibfnamefont {P.}~\bibnamefont
  {Lepage}},\ }\href {https://arxiv.org/abs/nucl-th/9706029} {\enquote
  {\bibinfo {title} {How to {R}enormalize the {S}chr\"odinger {E}quation},}\ }
  (\bibinfo {year} {1997}),\ \Eprint {http://arxiv.org/abs/nucl-th/9706029}
  {arXiv:nucl-th/9706029 [nucl-th]} \BibitemShut {NoStop}%
\bibitem [{\citenamefont {Calci}\ \emph {et~al.}(2016)\citenamefont {Calci},
  \citenamefont {Navr\'atil}, \citenamefont {Roth}, \citenamefont
  {Dohet-Eraly}, \citenamefont {Quaglioni},\ and\ \citenamefont
  {Hupin}}]{CNR16}%
  \BibitemOpen
  \bibfield  {author} {\bibinfo {author} {\bibfnamefont {A.}~\bibnamefont
  {Calci}}, \bibinfo {author} {\bibfnamefont {P.}~\bibnamefont {Navr\'atil}},
  \bibinfo {author} {\bibfnamefont {R.}~\bibnamefont {Roth}}, \bibinfo {author}
  {\bibfnamefont {J.}~\bibnamefont {Dohet-Eraly}}, \bibinfo {author}
  {\bibfnamefont {S.}~\bibnamefont {Quaglioni}}, \ and\ \bibinfo {author}
  {\bibfnamefont {G.}~\bibnamefont {Hupin}},\ }\href {\doibase
  10.1103/PhysRevLett.117.242501} {\bibfield  {journal} {\bibinfo  {journal}
  {Phys. Rev. Lett.}\ }\textbf {\bibinfo {volume} {117}},\ \bibinfo {pages}
  {242501} (\bibinfo {year} {2016})}\BibitemShut {NoStop}%
\bibitem [{\citenamefont {Baye}\ \emph {et~al.}(2005)\citenamefont {Baye},
  \citenamefont {Capel},\ and\ \citenamefont {Goldstein}}]{BCG05}%
  \BibitemOpen
  \bibfield  {author} {\bibinfo {author} {\bibfnamefont {D.}~\bibnamefont
  {Baye}}, \bibinfo {author} {\bibfnamefont {P.}~\bibnamefont {Capel}}, \ and\
  \bibinfo {author} {\bibfnamefont {G.}~\bibnamefont {Goldstein}},\ }\href
  {\doibase 10.1103/PhysRevLett.95.082502} {\bibfield  {journal} {\bibinfo
  {journal} {Phys. Rev. Lett.}\ }\textbf {\bibinfo {volume} {95}},\ \bibinfo
  {pages} {082502} (\bibinfo {year} {2005})}\BibitemShut {NoStop}%
\bibitem [{\citenamefont {Goldstein}\ \emph {et~al.}(2006)\citenamefont
  {Goldstein}, \citenamefont {Baye},\ and\ \citenamefont {Capel}}]{GBC06}%
  \BibitemOpen
  \bibfield  {author} {\bibinfo {author} {\bibfnamefont {G.}~\bibnamefont
  {Goldstein}}, \bibinfo {author} {\bibfnamefont {D.}~\bibnamefont {Baye}}, \
  and\ \bibinfo {author} {\bibfnamefont {P.}~\bibnamefont {Capel}},\ }\href
  {\doibase 10.1103/PhysRevC.73.024602} {\bibfield  {journal} {\bibinfo
  {journal} {Phys. Rev. C}\ }\textbf {\bibinfo {volume} {73}},\ \bibinfo
  {pages} {024602} (\bibinfo {year} {2006})}\BibitemShut {NoStop}%
\bibitem [{\citenamefont {Fukuda}\ \emph {et~al.}(2004)\citenamefont {Fukuda},
  \citenamefont {Nakamura}, \citenamefont {Aoi}, \citenamefont {Imai},
  \citenamefont {Ishihara}, \citenamefont {Kobayashi}, \citenamefont {Iwasaki},
  \citenamefont {Kubo}, \citenamefont {Mengoni}, \citenamefont {Notani},
  \citenamefont {Otsu}, \citenamefont {Sakurai}, \citenamefont {Shimoura},
  \citenamefont {Teranishi}, \citenamefont {Watanabe},\ and\ \citenamefont
  {Yoneda}}]{Fuk04}%
  \BibitemOpen
  \bibfield  {author} {\bibinfo {author} {\bibfnamefont {N.}~\bibnamefont
  {Fukuda}}, \bibinfo {author} {\bibfnamefont {T.}~\bibnamefont {Nakamura}},
  \bibinfo {author} {\bibfnamefont {N.}~\bibnamefont {Aoi}}, \bibinfo {author}
  {\bibfnamefont {N.}~\bibnamefont {Imai}}, \bibinfo {author} {\bibfnamefont
  {M.}~\bibnamefont {Ishihara}}, \bibinfo {author} {\bibfnamefont
  {T.}~\bibnamefont {Kobayashi}}, \bibinfo {author} {\bibfnamefont
  {H.}~\bibnamefont {Iwasaki}}, \bibinfo {author} {\bibfnamefont
  {T.}~\bibnamefont {Kubo}}, \bibinfo {author} {\bibfnamefont {A.}~\bibnamefont
  {Mengoni}}, \bibinfo {author} {\bibfnamefont {M.}~\bibnamefont {Notani}},
  \bibinfo {author} {\bibfnamefont {H.}~\bibnamefont {Otsu}}, \bibinfo {author}
  {\bibfnamefont {H.}~\bibnamefont {Sakurai}}, \bibinfo {author} {\bibfnamefont
  {S.}~\bibnamefont {Shimoura}}, \bibinfo {author} {\bibfnamefont
  {T.}~\bibnamefont {Teranishi}}, \bibinfo {author} {\bibfnamefont {Y.~X.}\
  \bibnamefont {Watanabe}}, \ and\ \bibinfo {author} {\bibfnamefont
  {K.}~\bibnamefont {Yoneda}},\ }\href {\doibase 10.1103/PhysRevC.70.054606}
  {\bibfield  {journal} {\bibinfo  {journal} {Phys. Rev. C}\ }\textbf {\bibinfo
  {volume} {70}},\ \bibinfo {pages} {054606} (\bibinfo {year}
  {2004})}\BibitemShut {NoStop}%
\bibitem [{\citenamefont {Moschini}\ and\ \citenamefont {Capel}(2019)}]{MC19}%
  \BibitemOpen
  \bibfield  {author} {\bibinfo {author} {\bibfnamefont {L.}~\bibnamefont
  {Moschini}}\ and\ \bibinfo {author} {\bibfnamefont {P.}~\bibnamefont
  {Capel}},\ }\href {\doibase https://doi.org/10.1016/j.physletb.2019.01.041}
  {\bibfield  {journal} {\bibinfo  {journal} {Phys. Lett.}\ }\textbf {\bibinfo
  {volume} {B790}},\ \bibinfo {pages} {367} (\bibinfo {year}
  {2019})}\BibitemShut {NoStop}%
\bibitem [{\citenamefont {Sparenberg}\ \emph {et~al.}(2010)\citenamefont
  {Sparenberg}, \citenamefont {Capel},\ and\ \citenamefont {Baye}}]{SCB10}%
  \BibitemOpen
  \bibfield  {author} {\bibinfo {author} {\bibfnamefont {J.-M.}\ \bibnamefont
  {Sparenberg}}, \bibinfo {author} {\bibfnamefont {P.}~\bibnamefont {Capel}}, \
  and\ \bibinfo {author} {\bibfnamefont {D.}~\bibnamefont {Baye}},\ }\href
  {\doibase 10.1103/PhysRevC.81.011601} {\bibfield  {journal} {\bibinfo
  {journal} {Phys. Rev. C}\ }\textbf {\bibinfo {volume} {81}},\ \bibinfo
  {pages} {011601} (\bibinfo {year} {2010})}\BibitemShut {NoStop}%
\bibitem [{\citenamefont {Pei}\ \emph {et~al.}(2006)\citenamefont {Pei},
  \citenamefont {Xu},\ and\ \citenamefont {Stevenson}}]{PXS06}%
  \BibitemOpen
  \bibfield  {author} {\bibinfo {author} {\bibfnamefont {J.}~\bibnamefont
  {Pei}}, \bibinfo {author} {\bibfnamefont {F.}~\bibnamefont {Xu}}, \ and\
  \bibinfo {author} {\bibfnamefont {P.}~\bibnamefont {Stevenson}},\ }\href
  {\doibase https://doi.org/10.1016/j.nuclphysa.2005.10.004} {\bibfield
  {journal} {\bibinfo  {journal} {Nucl. Phys.}\ }\textbf {\bibinfo {volume}
  {A765}},\ \bibinfo {pages} {29} (\bibinfo {year} {2006})}\BibitemShut
  {NoStop}%
\bibitem [{\citenamefont {Kelley}\ \emph {et~al.}(2012)\citenamefont {Kelley},
  \citenamefont {Kwan}, \citenamefont {Purcell}, \citenamefont {Sheu},\ and\
  \citenamefont {Weller}}]{KKP12}%
  \BibitemOpen
  \bibfield  {author} {\bibinfo {author} {\bibfnamefont {J.}~\bibnamefont
  {Kelley}}, \bibinfo {author} {\bibfnamefont {E.}~\bibnamefont {Kwan}},
  \bibinfo {author} {\bibfnamefont {J.}~\bibnamefont {Purcell}}, \bibinfo
  {author} {\bibfnamefont {C.}~\bibnamefont {Sheu}}, \ and\ \bibinfo {author}
  {\bibfnamefont {H.}~\bibnamefont {Weller}},\ }\href {\doibase
  https://doi.org/10.1016/j.nuclphysa.2012.01.010} {\bibfield  {journal}
  {\bibinfo  {journal} {Nucl. Phys.}\ }\textbf {\bibinfo {volume} {A880}},\
  \bibinfo {pages} {88} (\bibinfo {year} {2012})}\BibitemShut {NoStop}%
\bibitem [{\citenamefont {Capel}\ and\ \citenamefont {Nunes}(2006)}]{CN06}%
  \BibitemOpen
  \bibfield  {author} {\bibinfo {author} {\bibfnamefont {P.}~\bibnamefont
  {Capel}}\ and\ \bibinfo {author} {\bibfnamefont {F.~M.}\ \bibnamefont
  {Nunes}},\ }\href {\doibase 10.1103/PhysRevC.73.014615} {\bibfield  {journal}
  {\bibinfo  {journal} {Phys. Rev. C}\ }\textbf {\bibinfo {volume} {73}},\
  \bibinfo {pages} {014615} (\bibinfo {year} {2006})}\BibitemShut {NoStop}%
\bibitem [{\citenamefont {Capel}\ \emph {et~al.}(2004)\citenamefont {Capel},
  \citenamefont {Goldstein},\ and\ \citenamefont {Baye}}]{CGB04}%
  \BibitemOpen
  \bibfield  {author} {\bibinfo {author} {\bibfnamefont {P.}~\bibnamefont
  {Capel}}, \bibinfo {author} {\bibfnamefont {G.}~\bibnamefont {Goldstein}}, \
  and\ \bibinfo {author} {\bibfnamefont {D.}~\bibnamefont {Baye}},\ }\href
  {\doibase 10.1103/PhysRevC.70.064605} {\bibfield  {journal} {\bibinfo
  {journal} {Phys. Rev. C}\ }\textbf {\bibinfo {volume} {70}},\ \bibinfo
  {pages} {064605} (\bibinfo {year} {2004})}\BibitemShut {NoStop}%
\bibitem [{\citenamefont {Capel}\ \emph {et~al.}(2022)\citenamefont {Capel},
  \citenamefont {Phillips},\ and\ \citenamefont {Hammer}}]{CPH22}%
  \BibitemOpen
  \bibfield  {author} {\bibinfo {author} {\bibfnamefont {P.}~\bibnamefont
  {Capel}}, \bibinfo {author} {\bibfnamefont {D.}~\bibnamefont {Phillips}}, \
  and\ \bibinfo {author} {\bibfnamefont {H.-W.}\ \bibnamefont {Hammer}},\
  }\href {\doibase https://doi.org/10.1016/j.physletb.2021.136847} {\bibfield
  {journal} {\bibinfo  {journal} {Phys. Lett.}\ }\textbf {\bibinfo {volume}
  {B825}},\ \bibinfo {pages} {136847} (\bibinfo {year} {2022})}\BibitemShut
  {NoStop}%
\bibitem [{\citenamefont {Moro}\ and\ \citenamefont {Lay}(2012)}]{ML12}%
  \BibitemOpen
  \bibfield  {author} {\bibinfo {author} {\bibfnamefont {A.~M.}\ \bibnamefont
  {Moro}}\ and\ \bibinfo {author} {\bibfnamefont {J.~A.}\ \bibnamefont {Lay}},\
  }\href {\doibase 10.1103/PhysRevLett.109.232502} {\bibfield  {journal}
  {\bibinfo  {journal} {Phys. Rev. Lett.}\ }\textbf {\bibinfo {volume} {109}},\
  \bibinfo {pages} {232502} (\bibinfo {year} {2012})}\BibitemShut {NoStop}%
\bibitem [{\citenamefont {Kubushishi}\ and\ \citenamefont
  {Capel}(2024)}]{KC24}%
  \BibitemOpen
  \bibfield  {author} {\bibinfo {author} {\bibfnamefont {L.-P.}\ \bibnamefont
  {Kubushishi}}\ and\ \bibinfo {author} {\bibfnamefont {P.}~\bibnamefont
  {Capel}},\ }\href@noop {} {\enquote {\bibinfo {title} {Expanding {Halo-EFT}
  to core excitation for one-neutron halo nuclei: an application to
  $^{11}${Be}},}\ } (\bibinfo {year} {2024}),\ \bibinfo {note} {(In
  preparation)}\BibitemShut {NoStop}%
\end{thebibliography}%

\end{document}